\documentclass[12pt,preprint]{emulateapj}
\usepackage{amsfonts}
\usepackage{amssymb}
\usepackage{amsmath}

\def\Bpos {$\hbox{B}_{\rm pos}$}

\def\percc {$\mathrm{cm^{-3}}$} 
\def\kms {km s$^{-1}$}
\def\Msol {$M_\odot$}

\shorttitle{NGC~7538~IRS~1: Interaction of a polarized dust spiral and a molecular outflow}
\shortauthors{Wright et al.}

\begin{document}
\title{NGC~7538~IRS~1: Interaction of a polarized dust spiral and a molecular outflow}
\author{M. C. H. Wright\altaffilmark{1},
Charles~L.~H.~Hull\altaffilmark{1},
Thushara~Pillai\altaffilmark{2},
Jun-Hui Zhao\altaffilmark{3},
G\"oran Sandell\altaffilmark{4}
}
\altaffiltext{1}{Department of Astronomy, University of California, Berkeley, Berkeley, CA 94720, USA}
\altaffiltext{2}{Max Planck Institut f\"ur Radioastronomie, Auf dem H\"ugel 69, 53121 Bonn}
\altaffiltext{3}{Harvard-Smithsonian Center for Astrophysics, 60 
Garden Street, Cambridge, MA 02138, USA}
\altaffiltext{4}{SOFIA-USRA, NASA Ames Research Center,  MS
232-12, Building N232, Rm. 146, P.O. Box 1, Moffett Field, CA 94035-0001, USA}

\begin{abstract}

We present dust polarization and CO molecular line images of 
NGC~7538~IRS~1. We combined data from the SMA, CARMA and JCMT 
telescopes to make images with  $\sim$\,2.5$\arcsec$ resolution at  
230 and 345\,GHz. The images show a remarkable spiral pattern in both
the dust polarization and molecular outflow. These data dramatically 
illustrate the interplay between a high infall rate onto IRS~1 and
a powerful outflow disrupting the dense, clumpy medium surrounding 
the star. The images of the dust polarization
and the CO outflow presented here provide observational evidence for the exchange of 
energy and angular momentum between the infall and the outflow. 
The  spiral dust pattern, which rotates through over 180$^{\circ}$ from IRS~1, may be
a clumpy filament wound up by conservation of angular momentum in the infalling material.
The redshifted CO emission ridge traces the dust spiral closely through the MM dust cores, several of
which may contain protostars. We propose that the CO maps
 the boundary layer where the outflow is ablating gas from the dense gas in the spiral.

\end{abstract}

\section{Introduction}

NGC~7538~IRS~1 is a young, heavily 
accreting hyper-compact (HC) \ion{H}{2} region located at a distance of 2.65\,kpc \citep{mosc09}.
IRS~1 drives a well collimated bipolar ionized jet north-south (N--S)
in the central 0.025\,pc (2$\arcsec$) region \citep{camp84,gaum95,sand09}. 
With an IR luminosity of 10$^5$\,L$_\odot$ \citep{will76,akab05}, 
IRS~1 is also believed to be the energy source driving the CO ouflow, which extends  
NW--SE for 0.5\,pc (40$\arcsec$)  \citep[e.g.,][]{scov86,kame89,davi98,qiu11}. 
Single-dish data from the Five College Radio Astronomy Observatory (FCRAO), 
the Onsala Space Observatory (OSO), and the James Clerk Maxwell Telescope (JCMT), 
along with interferometric data from the Very Large Array (VLA), Berkeley-Illinois-Maryland Array (BIMA), 
Combined Array for Research in Millimeter-wave Astronomy (CARMA), and 
Submillimeter Array\footnote{The Submillimeter Array is a joint project between 
the Smithsonian Astrophysical Observatory and the Academia
Sinica Institute of Astronomy and Astrophysics and is funded by the
Smithsonian Institution and the Academia Sinica.} (SMA) show that the molecular outflow 
extends N--S at distances of $>$ 2\,pc (2.5\arcmin) \citep{sand12}.
Precession has been  suggested to explain why the inner ionized jet (0.025\,pc) and 
the molecular outflow are misaligned  { \citep{krau06}}. 

The spectral line profiles in $^{13}$CO(2--1), CO(2--1), and HCN(1--0) 
observed toward IRS\,1 show broad redshifted absorption, providing 
evidence for gas with an infall rate of 
$3-10\times10^{-3}$ M$_\odot$ yr$^{-1}$ \citep{zhu13}.
Observational evidence for an accretion disk is controversial.
A structure oriented NE--SW is observed
in the mid-infrared \citep{debu05}, as is  
a velocity gradient in the molecular lines \citep{brog08,klaa09,beut12,zhu13}. 
A NW--SE velocity gradient of $\sim$\,0.02\,km\,s$^{-1}$\,AU$^{-1}$  found in methanol masers \citep{mini98,mini00}
was modeled as a circumstellar disk \citep{pest04}. 
However, failing to find convincing evidence for an 
accretion disk from the sub-arcsec resolution observations 
with SMA and CARMA, \cite{zhu13} suggested that
a rotating ionized outflow could entrain the adjacent 
molecular gas, and the impact of the ionized outflow on the 
circumstellar gas could produce highly excited molecular 
species in the regions NE and SW of IRS~1. 

Recent dust polarization observations at 230\,GHz  \citep{Hull2014}
and 345\,GHz \citep{Frau2014} reveal a spiral distribution in both the dust emission
and the magnetic field (B-field) inferred from dust polarization.
Here we present dust polarization and CO molecular line 
images of NGC~7538~IRS~1. We combined data from the 
SMA, CARMA and 
JCMT telescopes to make images with  $\sim$\,2.5$\arcsec$
resolution at  230 and 345\,GHz. We discuss the possible origin of the 
remarkable spiral pattern observed in the molecular outflow, the dust emission, and the B-field.

\section{Observations and Data Reduction}

\subsection{SMA Data}

The SMA CO(3--2) line image was made from the SMA archival data 
observed on three consecutive days (2005 Oct 05, Oct 06, and Oct 07) 
 in a compact array configuration,
 and centered on IRS~1 at a position RA(J2000)\,=\,23:13:45.36,
Decl(J2000)\,=\,+61:28:10.6 in the NGC~7538 complex. The
observing frequency was set at $\nu_{\textrm{LO}}=341.5$\,GHz in a circular
polarization correlator mode with 24 adjacent spectral windows for
each sideband; each of the two sidebands contains 128 spectral channels with
widths of 0.812\,MHz. The upper sideband (USB) contains the
CO(3--2) line at $\nu_0=345.796$\,GHz. The data reduction was processed
in MIRIAD. The RR and LL visibility data were extracted prior to
calibrations. Mars and/or 3C454.3 were used for bandpass calibration.
The complex gains were calibrated using BL Lac and the flux-density
scale was determined using Uranus. The baseline-based corrections
were made using the point-source model determined from 3C111. The
residual errors in the gain were further corrected by using the
unresolved continuum source at IRS~1. The corrections for the residual
errors were applied to the CO line data in which the continuum level
was subtracted using the MIRIAD task UVLIN with a linear fitting.

The visibility channel data were binned to a velocity width of 1\,km s$^{-1}$
to make spectral line images with robust=2 weighting. The synthesized image
was cleaned in MIRIAD using the default mode. The typical RMS noise
is 0.35\,Jy beam$^{-1}$ per channel, with a synthesized beam FWHM of
$2.40\arcsec \times 2.05\arcsec$ (PA = 30.7$^{\circ}$).

\subsection{CARMA Data}

Observations were made with CARMA between May 2011 and April 2013.
Three different array configurations were used: C (26--370\,m baselines, or
telescope spacings), D (11--148\,m), and E (8.5--66\,m), which correspond to
angular resolutions at 1.3\,mm of approximately $1\arcsec$,
$2\arcsec$, and $4\arcsec$, respectively. These data were combined to make
images of the dust polarization and CO(2--1) emission with  $\sim$\,2.5$\arcsec$ resolution.
The observations of NGC~7538 are
part of the TADPOL survey. For details of these observations see \citet{Hull2014}.

\subsection{JCMT Data}


CO(3--2) observations were made with JCMT in service mode under program ID M05AI07 on
2005  Apr 14,  Apr 28, and May 20.
These data were combined with the SMA data using the MIRIAD task IMMERGE
to make well sampled images with angular resolution $\sim$\,2.5$\arcsec$.

\section{Results \label{results}}

\subsection{Dust polarization}

\begin{figure}
\includegraphics[width=10cm,angle=-90]{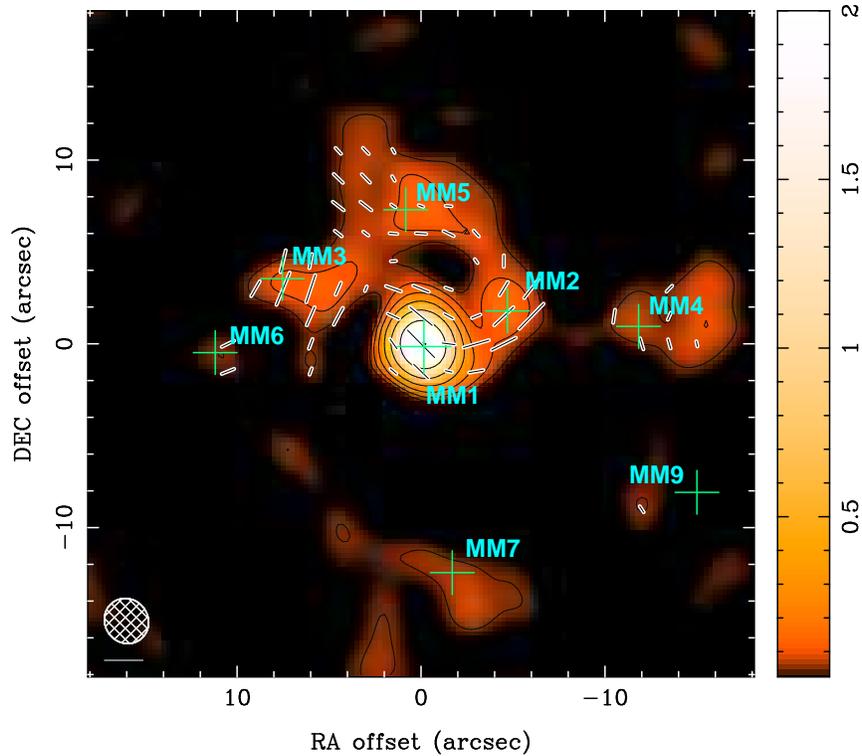} 
\centering
\caption{ { CARMA 1.3\,mm B-field vectors overlaid on 1.3\,mm dust continuum (colorscale and contours). 
Contours at 2,4,8,16,32,64,128 $\times$  25 mJy  beam$^{-1}$. The peak is 3.23 Jy  beam$^{-1}$. 
The RMS noise in the dust continuum image is 19.3  mJy  beam$^{-1}$.
The color wedge on the right shows the color scale with a logarithmic range from 0.025 to 2 Jy  beam$^{-1}$.
The positions are offsets from IRS~1 at  R.A. (J2000)\,=\,23$^{\rm h}$13$^{\rm m}$45.$^{\rm s}$37, 
Dec. (J2000)\,=\,61$\arcdeg$28$\arcmin$10\farcs43. 
The crosses mark the positions of SMA dust sources.
The  vectors of polarized intensity have been rotated by 90$^{\circ}$ to show the
POS B-field orientation.  
The synthesized beam FWHM is drawn in the bottom left corner.
The scale bar below the beam is drawn the length of the longest
vector of polarized intensity, 11.6 mJy  beam$^{-1}$. 
The RMS noise in the polarized intensity image is 0.7  mJy  beam$^{-1}$.}
\label{fig:polmap}}
\end{figure}

 Figure~\ref{fig:polmap} summarizes the results of the CARMA polarization observations at 230\,GHz (Hull et al. 2014).
Similar results were obtained by \citet{Frau2014} at 345\,GHz, confirming the remarkable spiral
structure.  The dust continuum emission is shown in 
color with the plane-of-the-sky (POS) component of the B-field (\Bpos) overlaid as vectors. 
Throughout the field of view (FOV) there is a high polarization fraction ($\sim$\,6\%) and remarkable alignment 
of the B-field vectors along a spiral structure centered on IRS~1. 
The only significant deviation of the B-field from the spiral pattern occurs near the dust clump MM5.  
The presence of many H$_2$O masers \citep{kame90} suggests that
stars may already have formed in some of the dust clumps, with outflows that have disrupted the spiral B-field pattern.
There is a marked decrease in the polarized 
fraction towards the center of IRS~1 ($\sim$\,1\%), as observed towards 
most star forming regions, where depolarization creates a ``hole'' 
due to the star formation process itself.

Outside of IRS~1, we use clumps MM2 and MM3 (following \citealt{qiu11}),
which have sufficient polarization for us to  estimate the strength of the B-field using the
Chandrasekhar-Fermi (CF) method \citep{chandra1953}. 

We can determine \Bpos\ towards 
MM2/MM3 using $B_{\rm pos}=f \sqrt{4\pi\rho} \, \delta V/\delta\phi$, 
where $\rho$ is the gas density, $\delta V$ is the velocity dispersion and
$\delta\phi$ is the dispersion in the polarization angle and $f$ is a
factor of $\sim\,0.5$ that corrects for any smoothing effect due to B-field 
averaging along the line of sight (LOS), provided the field is strong 
(${\delta\phi} < 25^{\circ}$) \citep{ostriker01}. Using the values from
Table\,\ref{tab:polzn}, we infer \Bpos\ for MM2 and MM3 
to be $\sim$\,5.6 and 7.5\,mG, respectively.\footnote{Note 
that the CF method assumes that magnetic energy is in equipartition with turbulent energy. If the B-field alignment along the spiral is induced either by the
outflow or infall, then the assumption of equipartition breaks down. Since the polarization angle due to turbulence is smoothed out
by the spiral pattern, the derived field strengths reported in Table \ref{tab:polzn} are only upper limits.}

A recently developed statistical
  approach to analyse the polarization angles allows us to
  estimate the contribution due to perturbations via the
  dispersion function, $1-\left \langle cos[\Delta \phi (l)] \right
  \rangle$, where $\Delta \phi (l)$ is the difference in polarization
angle between any pair of vectors separated by distance $l$
\citep{houde09}. We apply this method to our data and refer the reader
to \citet{Frau2014} for a discussion of the method  and its
specific application to the polarization data in NGC~7538.  The
345\,GHz data presented in Frau et al. are more sensitive to the dust emission and thus
allow them to treat the spiral and the central source (IRS~1)
separately. However, due to limited sensitivity in our 230\,GHz data,
we do not make this distinction. For the density and velocity dispersion
listed in Table\,\ref{tab:polzn}, and using our best fit value of the 
turbulent-to-large-scale magnetic field strength ratio of $0.59 \pm 0.03$, 
we derive a magnetic field strength of $\sim$\,2.2\,mG.
 This is consistent with the results from Frau et al.

\subsection{CO  Outflow}
 Figure~\ref{fig:CO32} shows  red- and blueshifted CO(3--2) outflows constructed
 from the image that combines SMA and JCMT data.
We compared CARMA CO(2--1) and SMA CO(3--2) channel images at the same
angular resolution.  The CO(2--1) structure observed with CARMA shows the same clumpy distribution, 
in good agreement with the CO(3--2) structure observed with the SMA.  The image
combining SMA and JCMT data includes large scale structures that are not sampled by the interferometers,
and reveals the connected spiral structure, which closely follows the B-field orientation and the dust emission.


\section{Discussion}

\subsection{Interaction between the Magnetic field, Spiral Structure \&  Outflow}

The small-scale B-field direction is inconsistent with the NW--SE
orientation of the large-scale B-fields from single-dish
observations (Figure 36, \citealt{Hull2014}). On smaller
scales the CO outflow, dust continuum and B-field morphology are consistent
with a dominant spiral structure. This is suggestive of a direct
interaction between the outflow, B-field and the accreting dense gas. 

How do the energetics of the various physical processes compare?
\cite{qiu11} have estimated the outflow energy to be $5 \times 10^{46}$\,erg.  
The magnetic energy corresponding to a B-field energy
density $B^2/{8 \pi}$ over a spherical volume of radius $\sim$\,0.03\,pc
(this number corresponds to the radii of the MM2 and MM3 regions; see
Table~\,\ref{tab:polzn}) is $\sim\,1 - 2 \times 10^{45}$\, erg.
The spiral structure occupies 5-10\% of the volume of the outflow, so that the
energy density of the B field is comparable to that in the outflow.
A similar conclusion was obtained by \citet{Frau2014} for the non-thermal energy.
Thus, the ram pressure exerted by the massive outflow may be sufficient to drag and compress
the magnetic field. However, the outflow may not be sufficiently powerful
to penetrate and disrupt the dense accreting gas;
we propose instead that the spiral structure may be deflecting the inner
(lower-velocity) part of the outflow and moulding it into a spiral
pattern along its boundary. 
Single-dish CO observations of IRS~1
covering a region a few arcmin in size (much larger than the region considered here)
indicate that the outflow resumes its original N--S orientation further
away from IRS~1 (Sandell et al. in prep.).

\subsection{ Infall and Outflow}

\begin{figure}
\centering

  \centering
  \includegraphics[width=10cm, angle=270, origin=c]{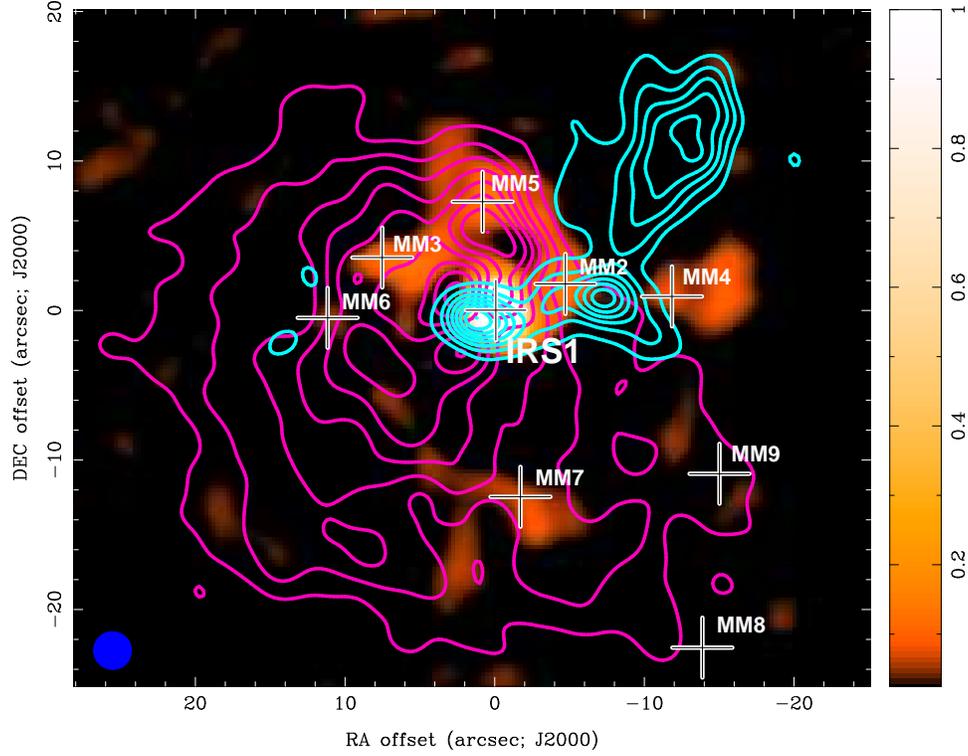}
  \vspace{0.25cm}
  \caption{ {Red- and blueshifted CO(3--2) emission overlaid on the dust polarization color image at 230\,GHz. 
   The dust continuum (color scale)  peak is 3.2 Jy  beam$^{-1}$.
   The color wedge on the right shows the color scale with a logarithmic range from 0.025 to 1 Jy  beam$^{-1}$.
  Magenta contours are in steps of 2.4 Jy  beam$^{-1}$ from 2.4 to 16.8  Jy  beam$^{-1}$ in velocity range 10  to 16 km s$^{-1}$ 
  from the systemic velocity --59.5 km s$^{-1}$.  
  Blue contours are in steps of 1.025 Jy  beam$^{-1}$ from 1.025 to 8.203 Jy  beam$^{-1}$ in velocity range --10  to --16 km s$^{-1}$ 
  from the systemic velocity --59.5 km s$^{-1}$. 
The positions are offsets from IRS~1 at  R.A. (J2000)\,=\,23$^{\rm h}$13$^{\rm m}$45.$^{\rm s}$37, 
Dec. (J2000)\,=\,61$\arcdeg$28$\arcmin$10\farcs43. 
The crosses mark the positions of SMA dust sources.
The synthesized beam FWHM is drawn in the bottom left corner.}
\label{fig:CO32}}
\end{figure}

At 1.3-mm wavelength, numerous molecular emission lines, including  optically thin, high
excitation energy levels which probe the innermost hot region (envelope/disk)  of IRS1
(e.g. OCS, CH$_3$CN, CH$_3$OCH$_3$ and C$_2$H$_5$OH),
show a strong emission peak at a radial velocity of $-59.5 \pm 1 $ km s$^{-1}$  \citep{zhu13}.
This value is close to the velocity component ($-59.7\pm0.3$ km~s$^{-1}$) observed
in the mid-infrared by \citet{knez09}.
Figure~\ref{fig:CO32} plots red- and blue-shifted CO 3-2 emission, more than 10 km s$^{-1}$
from a systemic velocity of --59.5 km s$^{-1}$ which we adopt in this paper.

 The red- and blueshifted outflow traced in  Figure~\ref{fig:CO32} closely follows the B-field orientation and the dust emission.  
 The redshifted CO emission shows a spiral structure starting from IRS~1,
 initially directed to the north, curving round at MM5 to the east, and continuing to change direction
 by more than 180$^{\circ}$. 
  The redshifted CO emission ridge traces the dust spiral closely through  MM 1, 2, 5, 3, 6, 7, and perhaps as far as MM9. 
 The blueshifted emission is initially directed to the SE within
 a few arcsec of IRS~1, then continues to the west, passing through a dust emission gap between
 MM2 and MM4 and extending toward the NW as observed in CO (1--0) on 100$\arcsec$ scales.

These results suggest substantial interaction between the outflow, and the spiral dust structure.
 \citet{Frau2014} detect 14 dust cores with masses 3.5 to 37\,M$_\odot$, and a total mass 160\,M$_\odot$, 
 in which star formation may be enhanced by ram pressure from the outflow.
 They model the velocity pattern observed in the spiral as expansion at 9\,km s$^{-1}$, and
 note that the outflow energy is comparable  to the spiral arm kinetic energy.
They suggest that the dust spiral is created from material swept up by the outflow.

 Although the model they propose Êsuggests that spiral dust feature and the magnetic field are 
tied to each other and are being pushed simultaneously by the outflow, 
it   does not take infall into account, and does not explain why the massive MM cores and the filament remain
aligned in a spiral pattern.

The infall and outflow have similar rates,  $\sim1- 3\times$10$^{-3}$ $\rm M_\odot$ yr$^{-1}$ 
 \citep{qiu11, zhu13}. 
The infall may have enough mass and momentum to deflect the outflow.
 An interesting alternative to Frau's expansion model is that the spiral
structure is a clumpy infalling filament, or the remains of a filament in which IRS~1 has formed, and is now
being ablated by the powerful outflow from IRS~1 (and possibly by other outflows in the region). 
The remarkable spiral structure, which rotates through over 180$^{\circ}$ from IRS~1, could be
wound up by conservation of angular momentum in the infalling material. 
A rotation of 1 km s$^{-1}$  would be  sufficient to wind the spiral structure through 180 degrees  in $\sim 4\times 10^5$ yr. Although, currently, there is an energetically dominant outflow, 
the mass and angular momentum ultimately originate in
the infall which created IRS1 and the compact MM sources.
The total mass in the MM sources and the filament, estimated from dust
emission  is  160\,M$_\odot$  \citep{Frau2014}.
The current infall rate is  $\sim1- 3\times$10$^{-3}$ $\rm M_\odot$ yr$^{-1}$ . An average infall
rate  $\sim4\times$10$^{-4}$ $\rm M_\odot$ yr$^{-1}$  would accumulate 160\,M$_\odot$  in $\sim 4\times 10^5$ yr.  
This is also the right timescale to wind up the filament.

For comparison,  \citet{qiu11} derive an outflow
mass 50  $\rm M_\odot$, energy  $4.9\times 10^{46}$  ergs,Ê dynamic time scale $2\times10^4$ yr, mass outflow
rate $2.5\times 10^{-3}$  $\rm M_\odot$ yr$^{-1}$, and momentum rate $2.3\times 10^{-2} $  $\rm M_\odot km~s^{-1} yr^{-1}$.
In our model, the CO is mapping material ablated from the filament
and reflects the radial velocity of the outflow, which is Frau et al's best
fit.
The observed CO (3-2) emission more than 10 km s$^{-1}$ from the systemic velocity  shown in Figure~\ref{fig:CO32}
maps material which has been swept up by the powerful outflow from IRS 1, and is consistent with the H$_{13}$CO$^+$ (4-3) and C$^{17}$O (3-2) mapped by  \citet{Frau2014} which also show the radial velocity of the outflow.

Submillimeter observations of NGC~7538 show filamentary dust ridges connecting IRS~1--3, NGC~7538~S, and IRS~9 \citep{sand04}. Hershel observations show that NGC~7538 has a highly filamentary
structure and multiple clumps with ongoing star formation \citep{Fallscheer2013}. These data
suggest that the spiral structure may be filamentary material in the envelope around IRS~1.

  Inverse P-Cygni profiles are seen towards IRS 1 in molecular lines with a wide range of excitation
 energy levels. Red-shifted absorption comes from dense gas in front of IRS 1 moving towards IRS 1. 
 Blue-shifted emission comes from gas behind IRS 1 moving towards the observer \citep{qiu11}. 
 Observations of HCO$^+$(1-0) by \cite{cord08} \& \cite{sand09} suggest that 
 the infall likely occurs within a region $\sim$ 0.06 pc. 
 Absorption features are detected in
 OCS(19-18) and CH$_3$CN(12-11) $k=$ 2 - 5 in the velocity range between $-56.5$ to
$-53.0$ km~s$^{-1}$, corresponding to  absorption features at
$-57.0$ and $-54.0$ km~s$^{-1}$ in HCN(1-0), CO(2-1) and $^{13}$CO(2-1)  \citep{zhu13}.
These absorptions are probably due to the redshifted absorption of
the infalling gas, revealing the hot dense gas in the accretion flow.
Absorption lines at higher excitation tracing warmer gas are more redshifted, perhaps tracing an accelerating
infall closer to IRS1. 
The  infall rate is estimated to be 1- 3$\times$10$^{-3}$ $\rm M_\odot$ yr$^{-1}$  \citep{qiu11, zhu13}.

\begin{figure}
  \centering
  \includegraphics[width=16cm, angle=0, origin=c]{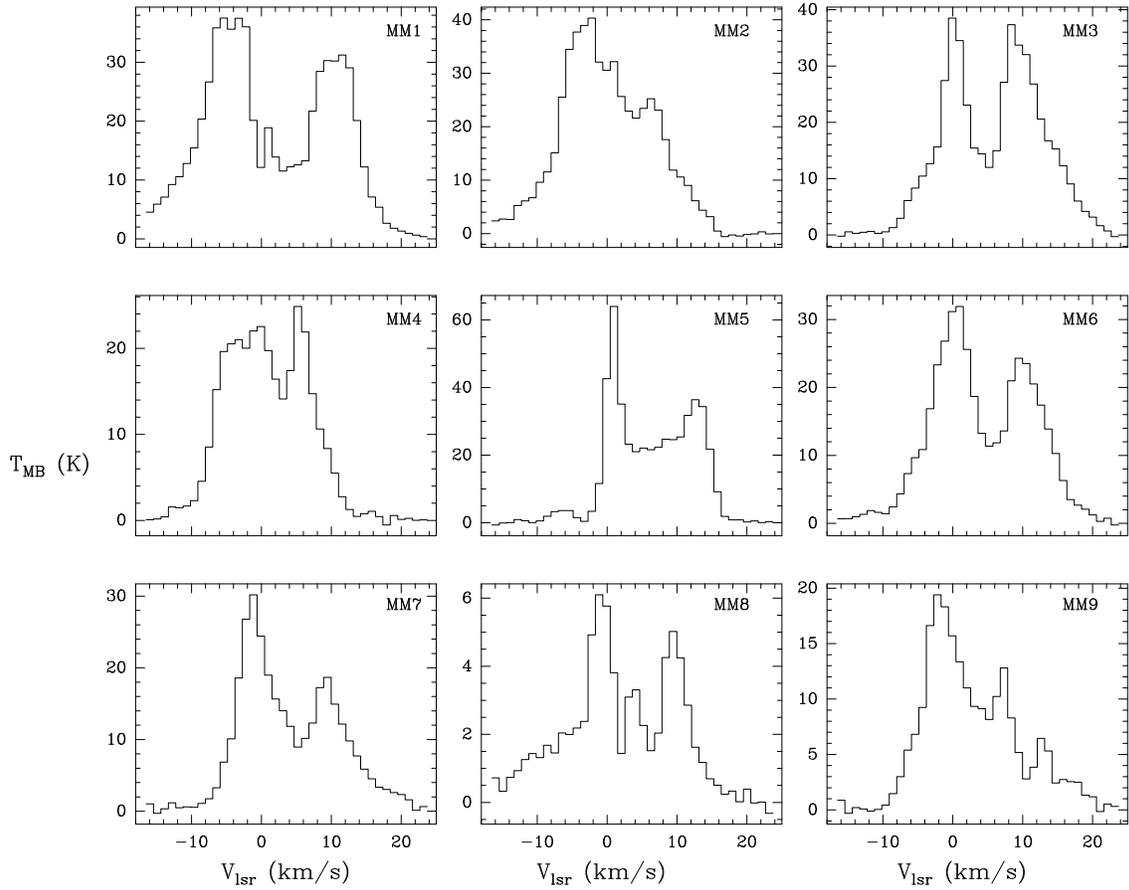}
  \vspace{0.25cm}
\caption{ CO(3--2) emission from --16 to +23\,km  s$^{-1}$ with respect to the systemic velocity --59.5\,km  s$^{-1}$. The CO(3--2) spectra at the positions of the MM sources marked in Figure 2.
Offset positions are with respect to  IRS~1 at R.A. (J2000) =
23$^{\rm h}$13$^{\rm m}$45.$^{\rm s}$37, Dec. (J2000) =
61$\arcdeg$28$\arcmin$10\farcs43.
\label{fig:CO32spectra}}
\end{figure}

Figure~\ref{fig:CO32spectra} shows the CO(3--2) spectra at the positions of the MM dust clumps. 
The spectra show multiple peaks corresponding to ambient and red- and blue-shifted
gas at most locations, suggesting a turbulent, wide angle outflow interacting with clumpy, infalling gas.
The  blue ``jet'' to the NW accelerates at increasing radius. Redshifted gas to the SE is more
 confined by higher densities. 
The CO(3--2) spectra are centered around +4 to +6\,km s$^{-1}$ with respect to the systemic velocity of $\sim$\,--59.5\,km s$^{-1}$.  The dip in the spectra around 
4 to 6\,km s$^{-1}$ may be due to absorption of the hot outflow by colder infalling gas at a velocity of $\sim$\,4 to 6\,km s$^{-1}$ with respect to the systemic velocity.
This is consistent with observations of the high-resolution
 absorption spectra of $^{13}$CO(2--1) and CO(2--1) \citep{zhu13}.
The  infalling envelope is  the ultimate source of the mass and angular momentum in the spiral.

  \citet{qiu11} trace multiple outflows with an estimated total mass outflow rate 
  2.5$\times$10$^{-3}$ $\rm M_\odot$ yr$^{-1}$ dominated by the outflow from IRS~1.

The molecular outflow mapped in
CO traces material that has been swept up from the ambient medium by the powerful outflow from
IRS~1.  Figure~\ref{fig:CO32} shows that the CO outflow follows the same spiral pattern traced by
the polarized dust emission, which contains dense star forming cores. 
We propose that the red- and blueshifted CO shown in Figure~\ref{fig:CO32} maps
 the boundary layer where the outflow is ablating gas from the dense gas in the spiral.

The images of dust polarization and CO outflow presented here provide observational 
evidence for the exchange of energy and angular
momentum between the infall and outflow.  This process is taking place throughout the infalling envelope as the outflow clears a way through the dense gas.  
In the paradigm model, a protostar creates an accretion disk
that then transports angular momentum away from, and mass towards, the forming star.  We have not seen the accretion disk, but we think it collimates the N--S ionized jet on subarcsec scales. 
A spiral structure extending into IRS1 could provide a ready explanation for the disparate position angles reported in
dense gas close to IRS1  \citep{brog08,klaa09,beut12,zhu13,mini98,mini00}.

But what happens further away from the protostar?
 Our observations suggest that the CO outflow is transporting
angular momentum from a dense spiral structure traced in polarized dust emission.
There could be significant angular momentum exchange 
in the infalling envelope, which would reduce the need to depend on a subarcsec accretion disk. 
A detailed mathematical model is beyond the scope of this paper,
but the outflow could capture angular momentum from the infall in successive interactions.
We may be seeing the process of an outflow transporting angular momentum from the infalling
gas away from the forming star.

\section{Conclusion}

A high accretion rate continues to fuel star formation around IRS~1, where both an early type O star
and a HC \ion{H}{2} region have formed. The ionized outflow from IRS~1 is
currently collimated in the N--S direction. The apparent rotation of the molecular outflow
and polarized dust emission suggests significant interaction between the outflow and accreting material. 
The infall and outflow have similar rates,  $\sim1- 3\times$10$^{-3}$ $\rm M_\odot$ yr$^{-1}$ .
The  spiral dust pattern, which rotates through over 180$^{\circ}$ from IRS~1,  may be
a clumpy filament wound up by conservation of angular momentum in the infalling material.
The redshifted CO emission ridge traces the dust spiral closely through the MM dust cores, several of
which may contain protostars. We propose that the CO maps
 the boundary layer where the outflow is ablating gas from the dense gas in the spiral.
 
 The outflow transports angular momentum away from
IRS~1 and may induce  gravitational instability in the massive clumps in the spiral pattern
forming protostars distributed around IRS~1.

\acknowledgments

C.L.H.H. acknowledges support from an NSF Graduate Fellowship and from a Ford
Foundation Dissertation Fellowship.  Support for CARMA
construction was derived from the states of California,
Illinois, and Maryland, the Gordon and Betty Moore
Foundation, the Kenneth T. and Eileen L. Norris Foundation,
the Associates of the California Institute of Technology,
and the National Science Foundation. Ongoing CARMA
development and operations are supported by the National
Science Foundation under a cooperative agreement, and by
the CARMA partner universities. 
The James Clerk Maxwell Telescope is operated by the Joint Astronomy Centre on behalf of the Science and Technology Facilities Council of the United Kingdom, the National Research Council of Canada, and (until 31 March 2013) the Netherlands Organisation for Scientific Research.
We thank the anonymous referee for a careful reading and some good questions and suggestions
which have improved this paper.

\begin{table}
\centering
\caption{Physical Parameters \label{tab:polzn}}
\begin{tabular}{ccccccccc}
\tableline
Clumps  & radius & $\delta V$  & Mass  & Density & $\delta\phi$ & \Bpos \\
 \tabularnewline
          & pc & \kms\ & \Msol\ &  10$^{6}$\,\percc & rad. & mG  \tabularnewline
\tableline 
MM2   & 0.027 & 1.7 & 12.2 & 2.1 & 0.17 & 5.6\tabularnewline
MM3   & 0.033 & 1.7 & 14.5 & 1.4 & 0.10 & 7.5  \tabularnewline
\tableline 
\end{tabular} \\
Notes: Mass and Density derived from CARMA 1.3\,mm dust continuum
assuming a dust temperature of 40\,K {over an effective area of
  radius $\rm R = \sqrt{(A/\pi)}$ \citep{sand04,qiu11}. We assume a
  dust opacity of 0.001 $\rm cm^2/g$ for dust grains with thin ice mantles and gas density $\rm n(H)=
10^6~$\percc\ as in \citet{ossenkopf1994:opacities}, and gas to dust
ratio of 100.  The velocity
  dispersion of 1.7\,\kms\ corresponds to the maximum linewidth
  reported in \citet{qiu11} based on Gaussian fits to different
  spectral lines at 230\,GHz.}

\end{table}


\end{document}